\documentclass[twocolumn,pra,preprintnumbers, showpacs, nofootinbib, superscriptaddress, amsmath,amssymb]{revtex4}
\usepackage{amstext,mathrsfs,amsthm,bm}
\usepackage{mathtools} 
\usepackage{stmaryrd} 

\begin{document}

\author{Denys I. Bondar}
\email{dbondar@princeton.edu}
\affiliation{Department of Chemistry, Princeton University, Princeton, NJ 08544, USA}

\author{Renan Cabrera}
\email{rcabrera@princeton.edu}
\affiliation{Department of Chemistry, Princeton University, Princeton, NJ 08544, USA} 

\author{Dmitry V. Zhdanov}
\email{dmitry.zhdanov@northwestern.edu}
\affiliation{Present address: Department of Chemistry, Northwestern University, Evanston, IL 60208, USA}
\affiliation{Department of Chemistry, Princeton University, Princeton, NJ 08544, USA} 

\author{Herschel A. Rabitz}
\email{hrabitz@princeton.edu}
\affiliation{Department of Chemistry, Princeton University, Princeton, NJ 08544, USA} 

\title{Wigner phase space distribution as a wave function}

\begin{abstract}
We demonstrate that the Wigner function of a pure quantum state is a wave function in a specially tuned Dirac bra-ket formalism and argue that the Wigner function is in fact a probability amplitude for the quantum particle to be at a certain point of the classical phase space. Additionally, we establish that in the classical limit, the Wigner function transforms into a classical Koopman-von Neumann wave function rather than into a classical probability distribution. Since probability amplitude need not be positive, our findings provide an alternative outlook on the Wigner function's negativity.
\end{abstract}

\date{\today}


\maketitle

\section{Introduction}

In his seminal work \cite{Wigner1932}, Wigner defined the combined distribution of the quantum particle's coordinate and momentum in terms of the wave function. Since then, the Wigner function has played a paramount role in the phase space formulation of quantum mechanics \cite{Bolivar2004, Zachos2005, Curtright2011}, is a standard tool for establishing the quantum-to-classical interface \cite{Zurek2001, Bolivar2004, dragoman2004quantum, Haroche2006}, and has a broad range of applications in optics and signal processing \cite{Cohen1989, Dragoman2005} as well as quantum computing \cite{Miquel2002a, Galvao2005, Cormick2006, Ferrie2009a, Veitch2012a, Mari2012, Veitch2013}.  Techniques for the experimental measurement of the Wigner function are also developed \cite{Kurtsiefer1997, Haroche2006, Ourjoumtsev2007, Deleglise2008}. Despite its ubiquity, the Wigner function is haunted by the obscure feature of possibly being negative. Wigner \cite{O'Connell1981} demonstrated that his function is the only one satisfying a reasonable set of axioms for a joint probability distribution. This feature of the Wigner function has been a subject of numerous interpretations \cite{Bracken2004, Man'ko2004, Ferrie2009a, Ferrie2011}, including the development of a mathematical framework for handling negative probabilities \cite{Muckenheim1986, Landsberg1987}. The Wigner function's negativity was also associated with the exponential speedup in quantum computation \cite{Veitch2012a, Veitch2013}.

In this paper we provide insight into the negativity by advocating the following interpretation: The Wigner function is a probability amplitude for a quantum particle to be at a certain point of the classical phase space, i.e., the Wigner function is a wave function analogous to the Koopman von-Neumann (KvN) wave function of a classical particle.

The remainder of the paper is organized as follows: The necessary background on the KvN classical mechanics and operational dynamical modeling \cite{Bondar2011c} is reviewed in Sec. \ref{Sec:Background_ODM}. The basic equations, on which our interpretation rests, are derived in Secs. \ref{Sec:Derivation_Hilber_phase_space}-\ref{Sec:Hilbert_phase_space_realization}. A connection between these equations and quantum mechanics in phase space is established in Sec. \ref{Sec:Commenst_on_phase_space_formalizm}. Conclusions are drawn in Sec. \ref{Sec:Conclusions}.

\section{Background}\label{Sec:Background_ODM}

Around the time the Wigner distribution was conceived, Koopman and von Neumann \cite{Koopman1931, Neumann1932, Neumann1932a, DaniloMauro2002} (for modern developments and applications see Refs. \cite{Gozzi1988, Gozzi1989, Wilkie1997, Wilkie1997a, Gozzi2002, DaniloMauro2002, Deotto2003, Deotto2003a, Abrikosovjr2005, Blasone2005, Brumer2006, Carta2006, Gozzi2010, Gozzi2011, Cattaruzza2011}) recast classical mechanics in a form similar to quantum mechanics by introducing classical complex valued wave functions and representing associated physical observables by means of commuting self-adjoint operators. In particular, it was postulated that the wave function $| \Psi(t) \rangle$ of a classical particle obeys the following equation of motion:
\begin{align}
	& i\frac{d}{dt} | \Psi(t) \rangle = \hat{L} | \Psi(t) \rangle, \quad \hat{L} = \frac{\hat{p}}{m} \hat{\lambda}_x  - U'(\hat{x}) \hat{\lambda}_p, 
		\label{Ch2_AlmostLiouville_Eq} \\
	& [ \hat{x}, \hat{\lambda}_x ] = [ \hat{p}, \hat{\lambda}_p ] = i, \notag\\
	& [ \hat{x}, \hat{p} ] = [ \hat{x}, \hat{\lambda}_p ] =  [ \hat{p}, \hat{\lambda}_x ] = [ \hat{\lambda}_x, \hat{\lambda}_p ] = 0.
	\label{Complete_classical_algebra}
\end{align}	
Without loss of generality one-dimensional systems are considered throughout. Since the self-adjoint operators representing the classical observables of coordinate $\hat{x}$ and momentum $\hat{p}$ commute, they share a common set of orthogonal eigenvectors $| p \, x \rangle$ such that $1 = \int dpdx \, | p \, x \rangle\langle p \, x |$. In the KvN classical mechanics, all  observables are  functions of  $\hat{x}$ and $\hat{p}$. The expectation value of an observable $\hat{F} = F(\hat{x}, \hat{p})$ at time $t$ is $\langle \Psi(t) | \hat{F} | \Psi(t) \rangle$. The probability amplitude $\langle p \, x | \Psi(t) \rangle$ for a classical particle to be at point $x$ with momentum $p$ at time $t$ is found to satisfy
\begin{align}\label{PreLiouvillianEq}
		\left[ \frac{\partial }{\partial t} + \frac{p}{m} \frac{\partial}{\partial x} -  U'(x) \frac{\partial}{\partial p} \right]  \langle p \, x |\Psi(t) \rangle = 0.
\end{align}
This is the evolution equation for the classical wave function in the $xp$-representation, where $\hat{x} = x$, $\hat{\lambda}_x = -i\partial / \partial x$, $\hat{p} = p$, and $\hat{\lambda}_p = -i \partial / \partial p$ in order to satisfy the commutation relations (\ref{Complete_classical_algebra}). Utilizing the chain rule and equation (\ref{PreLiouvillianEq}), we obtain the well known classical Liouville equation for the phase space probability distribution $\rho(x,p;t) = |\langle p \, x | \Psi(t) \rangle |^2$,
\begin{align}\label{Liouville_Eq}
	\left[ \frac{\partial }{\partial t} + \frac{p}{m} \frac{\partial}{\partial x} -  U'(x) \frac{\partial}{\partial p}  \right]  \rho(x,p;t) = 0.
\end{align}

\begin{figure}
	\begin{center}
		\includegraphics[scale=0.33]{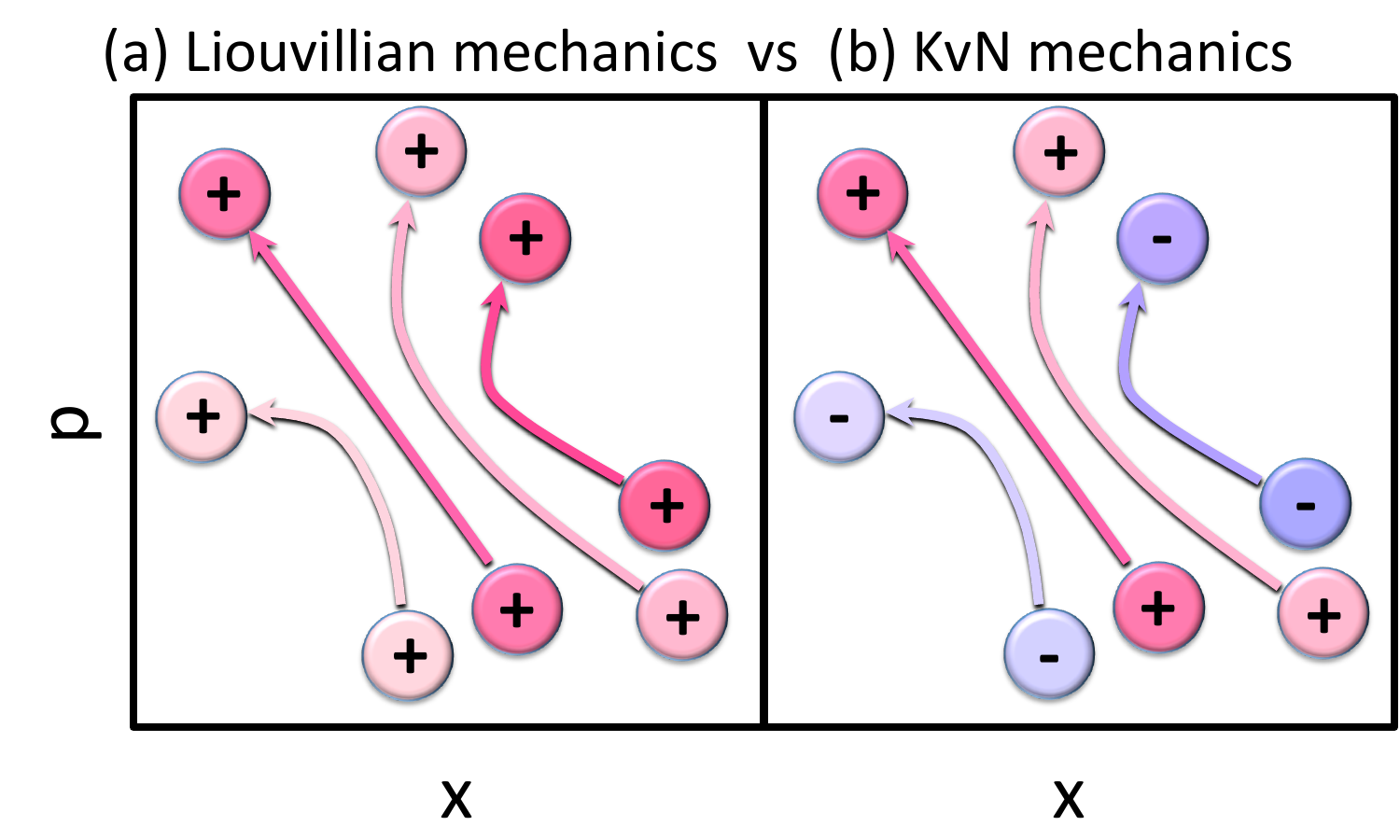}
		\caption{(color online) The conceptual difference between (a) Liouville and (b) KvN classical mechanics. In the Liouville mechanics [equation (\ref{Liouville_Eq})], classical particles, moving along Newtonian trajectories, are tagged by positive weights according to the probability distribution $\rho$. In the KvN mechanics [equation (\ref{PreLiouvillianEq})], classical particles, following the same  trajectories, are tagged by arbitrary real (or complex) weights representing the KvN classical wave function $| \Psi \rangle$.}
			\label{Fig_Liuville_vs_KvN}
	\end{center}
\end{figure}

Newtonian trajectories emerge as characteristics of either equation (\ref{PreLiouvillianEq}) or (\ref{Liouville_Eq}). Hence, the essential difference between KvN and Liouville mechanics lies in weighting individual trajectories (see figure \ref{Fig_Liuville_vs_KvN}): Arbitrary complex weights underlying the classical wave function $| \Psi \rangle$ can be utilized in KvN mechanics (\ref{PreLiouvillianEq}); whereas, only positive weights having probabilistic meaning are permitted in Liouville mechanics (\ref{Liouville_Eq}). 

Note that the classical wave function and the classical probability distribution satisfy the same dynamical equation, which reflects the physical irrelevance of the phase of a classical wave function. 

\begin{figure}
	\begin{center}
		\includegraphics[scale=0.30]{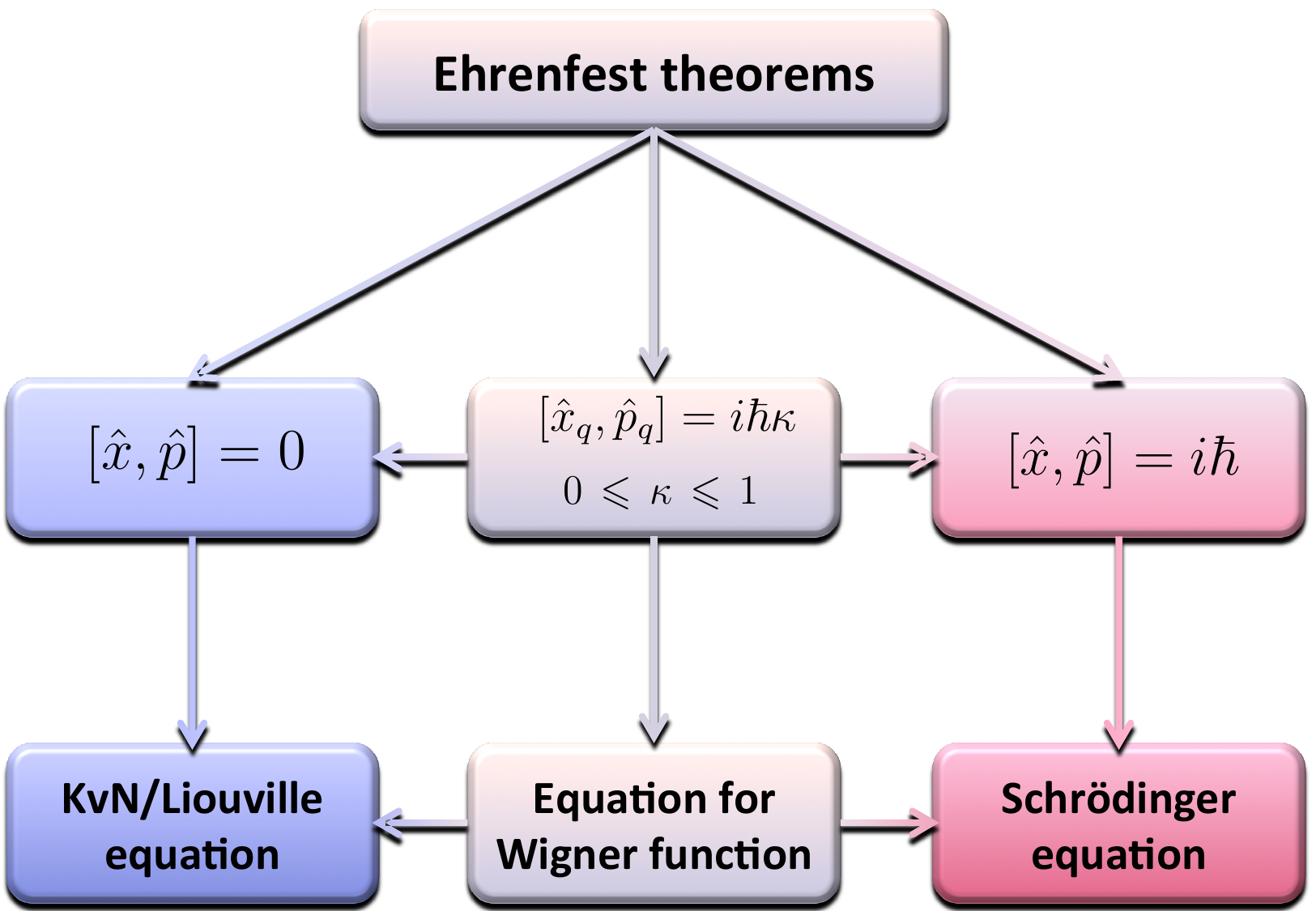}
		\caption{(color online) Schematic flow showing the derivation of quantum mechanics, classical mechanics, and quantum mechanical phase space representation within operational dynamical modeling  proposed in \cite{Bondar2011c}.}
			\label{Fig_EhrenfestQuantization}
	\end{center}
\end{figure}

In recent work \cite{Bondar2011c}, we put forth operational dynamical modeling as a systematic theoretical framework for deducing equations of motion from the evolution of average values. First, starting from the Ehrenfest theorems \cite{Ehrenfest1927}, we obtained the Schr\"{o}dinger equation if the momentum and coordinate operators obeyed the canonical commutation relation, and the KvN equation (\ref{Ch2_AlmostLiouville_Eq}) if the momentum and coordinate operators commuted. Then, applying the same technique to the Ehrenfest theorems,
\begin{align}\label{EhrenfestThrs_in_quantumclass_case}
	m\frac{d}{dt} \langle \Psi_{\kappa}(t) | \hat{x}_q | \Psi_{\kappa}(t) \rangle &= \langle \Psi_{\kappa}(t) | \hat{p}_q | \Psi_{\kappa}(t) \rangle, \notag\\
	\frac{d}{dt} \langle \Psi_{\kappa}(t) | \hat{p}_q | \Psi_{\kappa}(t) \rangle &= \langle \Psi_{\kappa}(t)| -U'(\hat{x}_q) | \Psi_{\kappa}(t) \rangle,
\end{align}
with a generalization $[ \hat{x}_q, \hat{p}_q ] = i\hbar\kappa$ ($0 \leqslant \kappa \leqslant 1$) and demanding a smooth classical limit $\kappa \to 0$, we established the existence of the uniquely defined operator $\hat{\mathcal{H}}_{qc}$ such that
\begin{align}
	& i \hbar \frac{d}{dt} | \Psi_{\kappa}(t) \rangle = \hat{\mathcal{H}}_{qc} | \Psi_{\kappa}(t) \rangle, \notag\\
	& \hat{\mathcal{H}}_{qc} =  \frac{\hbar}{m} \hat{p} \hat{\lambda}_x + 
						\frac{1}{\kappa} U\left( \hat{x} - \frac{\hbar\kappa}{2} \hat{\lambda}_p \right) -
						\frac{1}{\kappa} U\left( \hat{x} + \frac{\hbar\kappa}{2} \hat{\lambda}_p \right), \notag\\
	& \hat{x}_q = \hat{x} - \hbar\kappa \hat{\lambda}_p / 2, \qquad \hat{p}_q = \hat{p} + \hbar\kappa \hat{\lambda}_x / 2,
					\label{Dynamical_Eq_for_Wigner}
\end{align}
where $\hat{x}_q $ and $\hat{p}_q$ represent the quantum coordinate and momentum respectively, $\hat{x}$, $\hat{p}$, $\hat{\lambda}_x$, and $\hat{\lambda}_p$ are the same classical operators as in equation (\ref{Complete_classical_algebra}), and $\kappa$ denotes the degree of quantumness or commutativity: $\kappa \to 1$ corresponds to quantum mechanics whereas $\kappa \to 0$ recovers classical mechanics, and $\lim_{\kappa \to 0} \hat{\mathcal{H}}_{qc} = \hbar \hat{L}$.
See figure \ref{Fig_EhrenfestQuantization} for a pictorial summary of these derivations presented in \cite{Bondar2011c}.

A crucial point for our current analysis is that this unified wave function $| \Psi_{\kappa} \rangle$ ($t$ is dropped henceforth) in the $xp$-representation is proportional to the Wigner function $W$ (see Ref. \cite{Bondar2011c}),
\begin{align}
	& \langle p \, x | \Psi_{\kappa} \rangle =  \sqrt{2\pi \hbar\kappa} W(x, p), \notag\\
	& W(x, p) = \int \frac{dy}{2\pi \hbar\kappa} 
		\rho_{\kappa} \left(x - \frac{y}{2},  x + \frac{y}{2} \right) e^{ipy/(\hbar\kappa)};
		\label{UnifiedWaveFunc_connected_Wigner}
\end{align}
moreover,
\begin{align} \label{DenistyMatrixInHxp}
	& \langle x \lambda_p | \Psi_{\kappa} \rangle = \sqrt{\hbar\kappa} \rho_{\kappa}(u,v), \notag\\
	& \left[ i \hbar\kappa \frac{\partial}{\partial t} - \frac{(\hbar\kappa)^2}{2m}\left( \frac{\partial^2}{\partial v^2} 
	- \frac{\partial^2}{\partial u^2} \right) - U(u) + U(v) \right] \rho_{\kappa}(u,v)= 0, \notag\\
	& u = x - \hbar\kappa \lambda_p /2 , \qquad v = x + \hbar\kappa \lambda_p / 2. 
\end{align}

\section{Derivation of Main results}\label{Sec:Derivation_Hilber_phase_space}

Since $\hat{x}$ and $\hat{\lambda}_p$ are commuting self-adjoint operators, the following resolution of the identity is valid  in {\it the Hilbert phase space} $\mathcal{H}_{xp}$ (i.e., a vector space with the scalar product $\langle \quad | \quad \rangle $ of functions of two variables):
\begin{align}\label{LambdaPX_resulotion_identity}
	& \hat{1}_{\mathcal{H}_{xp}} = \int dx d\lambda_p \, | x \lambda_p \rangle\langle x \lambda_p|, 
		\qquad  | x \lambda_p \rangle \in \mathcal{H}_{xp}, \notag\\
	& \hat{x} | x \lambda_p \rangle = x | x \lambda_p \rangle, \qquad 
		\hat{\lambda}_p | x \lambda_p \rangle = \lambda_p | x \lambda_p \rangle;
\end{align}
likewise,
\begin{align}\label{LambdaXP_resulotion_identity}
	&\hat{1}_{\mathcal{H}_{xp}} = \int d\lambda_x dp \, | \lambda_x p \rangle\langle \lambda_x p |, \qquad
	 	 | \lambda_x p \rangle \in \mathcal{H}_{xp}, \notag\\
	& \hat{p} | \lambda_x p \rangle = p | \lambda_x p \rangle, 
		\qquad \hat{\lambda}_x | \lambda_x p \rangle = \lambda_x | \lambda_x p \rangle.
\end{align}
A consequence of the commutators $[\hat{x}, \hat{\lambda}_x] = [\hat{p}, \hat{\lambda}_p] = i$ is
\begin{align}
	\langle \lambda_x p | x \lambda_p \rangle = e^{ip\lambda_p - ix\lambda_x} / (2\pi). \notag
\end{align}

Let $\mathcal{H}_q$ denote the Hilbert space of single particle wave functions (i.e., a vector space with the scalar product $\langle \langle \quad || \quad \rangle\rangle$ of functions of one variable), which differs from the Hilbert phase space $\mathcal{H}_{xp}$ defined above. We denote by $\hat{x}_q$ and $\hat{p}_q$ ($[\hat{x}_q, \hat{p}_q]=i\hbar\kappa$) two pairs of operators: one acting on $\mathcal{H}_{xp}$ [as defined in Eq. (\ref{Dynamical_Eq_for_Wigner})], the other acting on $\mathcal{H}_q$, 
\begin{align}\label{XandPeigenvectsinHq}
	& \hat{x}_q || x_q \rangle\rangle =  x_q || x_q \rangle\rangle, \qquad \hat{p}_q || p_q \rangle\rangle =  p_q || p_q \rangle\rangle, \notag\\
	& \hat{1}_{\mathcal{H}_{q}} = \int dx_q || x_q \rangle\rangle \langle\langle x_q || = \int dp_q || p_q \rangle\rangle \langle\langle p_q ||, \notag\\ 
	& \langle\langle x_q || p_q \rangle\rangle = \frac{e^{ix_q p_q / (\hbar\kappa)}}{\sqrt{2\pi\hbar\kappa}},
	 \quad || x_q \rangle\rangle \in \mathcal{H}_{q}, \quad  || p_q \rangle\rangle \in \mathcal{H}_{q}.
\end{align}

The density matrix $\hat{\rho}_{\kappa}$, a self-adjoint operator acting in the space $\mathcal{H}_q$, obeys the Liouville -- von Neuman equation
\begin{align}
	i\hbar \kappa \frac{d}{dt} \hat{\rho}_{\kappa} = \left[ \frac{\hat{p}_q^2}{2m} + U(\hat{x}_q), \, \hat{\rho}_{\kappa}\right]\Longrightarrow \label{LiovillevonNeumanEq} \\
	\left[ i \hbar\kappa \frac{\partial}{\partial t} - \frac{(\hbar\kappa)^2}{2m}\left( \frac{\partial^2}{\partial v^2} 
	- \frac{\partial^2}{\partial u^2} \right) - U(u) + U(v) \right] \notag\\
	\times \langle\langle x_q = u || \hat{\rho}_{\kappa} || x_q' = v \rangle\rangle = 0. \label{DenistyMatrixInHq}
\end{align}
Comparing Eqs. (\ref{DenistyMatrixInHxp}) and (\ref{DenistyMatrixInHq}), we conclude that
\begin{align}\label{Density_matrix_coord_represent}
	\langle x \lambda_p | \Psi_{\kappa} \rangle = \sqrt{\hbar\kappa}  \langle\langle x_q = u || \hat{\rho}_{\kappa} || x_q' = v \rangle\rangle.
\end{align} 

Utilizing Eqs. (\ref{LambdaPX_resulotion_identity}), (\ref{XandPeigenvectsinHq}), and (\ref{Density_matrix_coord_represent}), we obtain
\begin{align}
	\langle \Psi_{\kappa} | \Psi_{\kappa} \rangle &= \int dx d\lambda_p \, \langle \Psi_{\kappa} | x \lambda_p \rangle \langle x \lambda_p | \Psi_{\kappa} \rangle \notag\\
	& = \hbar\kappa \int dx d\lambda_p \, \rho_{\kappa}^*(u,v) \rho_{\kappa}(u,v)  \notag\\
	&=  \int dx_q dx_q' \, \langle\langle x_q' || \hat{\rho}_{\kappa} || x_q \rangle\rangle \langle\langle x_q || \hat{\rho}_{\kappa} || x_q' \rangle\rangle \notag\\
	& = \int dx_q' \, \langle\langle x_q' || \hat{\rho}_{\kappa}^2   || x_q' \rangle\rangle = {\rm Tr}\left(\hat{\rho}_{\kappa}^2\right);
\end{align}
whence,
\begin{align}
	\langle \Psi_{\kappa} | \Psi_{\kappa} \rangle = 1 \Longleftrightarrow \hat{\rho}_{\kappa}^2 = \hat{\rho}_{\kappa}.
		\label{Normalization_of_Psi_kappa}
\end{align}

Setting $\hat{\rho}_{\kappa} =|| \phi_{\kappa} \rangle\rangle \langle\langle \phi_{\kappa} ||$, $\langle\langle \phi_{\kappa} || \phi_{\kappa} \rangle\rangle =1$, where $|| \phi_{\kappa} \rangle\rangle \in \mathcal{H}_{q}$ is a one-particle wave function, and denoting $\phi_{\kappa}(u) = \langle\langle x_q = u || \phi_{\kappa} \rangle\rangle$, $\phi_{\kappa}(v) = \langle\langle x_q = v || \phi_{\kappa} \rangle\rangle$, $\eta = p + \hbar\kappa\lambda_x /2$, and $\xi = p - \hbar\kappa\lambda_x /2$, we derive
\begin{widetext}
\begin{align}\label{Derivation_Equivalence_of_Averages}
	\langle \Psi_{\kappa}| G(\hat{x}_q) F(\hat{p}_q) | \Psi_{\kappa} \rangle 
	&= \int dx d\lambda_p dx' d\lambda_p' d\lambda_x dp \, \langle \Psi_{\kappa}| x \lambda_p \rangle \langle x \lambda_p | G(\hat{x}_q) F(\hat{p}_q) | \lambda_x p \rangle\langle \lambda_x p | x' \lambda_p' \rangle \langle x' \lambda_p' | \Psi_{\kappa} \rangle \notag\\
	&= \hbar\kappa \int dx d\lambda_p dx' d\lambda_p' d\lambda_x dp \, \phi_{\kappa}^* (u) \phi_{\kappa}(v) G(u) \langle x \lambda_p | \lambda_x p \rangle F(\eta) \langle \lambda_x p | x' \lambda_p' \rangle \phi_{\kappa} (u') \phi_{\kappa}^* (v') \notag\\
	&= \int \frac{du dv du' dv' d\xi d\eta}{(2\pi\hbar\kappa)^2} \phi_{\kappa}^* (u) \phi_{\kappa}(v) G(u) F(\eta) \phi_{\kappa} (u') \phi_{\kappa}^* (v') e^{i\eta(u-u')/(\hbar\kappa)} e^{-i\xi(v-v')/(\hbar\kappa)} \notag\\
	&= \left[ \int \phi_{\kappa}(v) \phi_{\kappa}^*(v)dv \right] \int d\eta \left[ \int \phi_{\kappa}^* (u) G(u) \frac{ e^{i\eta u / (\hbar\kappa)} }{\sqrt{2\pi\hbar\kappa}}du \right] F(\eta) \left[ \int \phi_{\kappa} (u') \frac{e^{-i\eta u' / (\hbar\kappa)}}{\sqrt{2\pi\hbar\kappa}} du' \right] \notag\\
	&= \int dp_q \left[ \int \phi_{\kappa}^* (x_q) G(x_q) \frac{ e^{ip_q x_q / (\hbar\kappa)} }{\sqrt{2\pi\hbar\kappa}}dx_q \right] F(p_q) \left[ \int \phi_{\kappa} (x_q') \frac{e^{-ip_q x_q' / (\hbar\kappa)}}{\sqrt{2\pi\hbar\kappa}} dx_q' \right] \notag\\
	&= \int dx_q dx_q' dp_q \, \langle\langle \phi_{\kappa} || G(\hat{x}_q) || x_q \rangle\rangle \langle\langle x_q || p_q \rangle\rangle \langle\langle p_q || F(\hat{p}_q) || x_q' \rangle\rangle \langle\langle x_q' || \phi_{\kappa} \rangle\rangle.
\end{align}
\end{widetext}
Finally, we have demonstrated that all expectation values for the state $|| \phi_{\kappa} \rangle\rangle$ coincide with those for $|\Psi_{\kappa} \rangle$,
\begin{align}
	\langle \Psi_{\kappa} | G(\hat{x}_q) F(\hat{p}_q)  | \Psi_{\kappa} \rangle =  \langle\langle \phi_{\kappa} ||  G(\hat{x}_q) F(\hat{p}_q) || \phi_{\kappa}  \rangle\rangle, \label{Equivalence_of_Averages}
\end{align}
where $G(\hat{x}_q)$ and $F(\hat{p}_q)$ are arbitrary functions of the quantum position and momentum, respectively.

Furthermore, in the context of the Maslov noncommutative calculus \cite{Maslov1976, Nazaikinskii1992, Nazaikinskii1996} based on the most general operator ordering, identity (\ref{Equivalence_of_Averages}) can be generalized to
\begin{align}
	\langle \Psi_{\kappa}| F(\hat{x}_q, \hat{p}_q) | \Psi_{\kappa} \rangle = \langle\langle \phi_{\kappa} || F(\hat{x}_q, \hat{p}_q) || \phi_{\kappa} \rangle\rangle, 
\end{align}
valid for an arbitrary function $F$ of two variables. 

Equations (\ref{Normalization_of_Psi_kappa}) and (\ref{Equivalence_of_Averages}) reveal that the Wigner distribution of a pure state behaves like a wave function. As shown in figure \ref{Fig_EhrenfestQuantization}, the Wigner function's dynamical equation (\ref{Dynamical_Eq_for_Wigner}) transforms into the evolution equation for a classical KvN wave function (\ref{Ch2_AlmostLiouville_Eq}). Hence, in the classical limit, the Wigner function maps a quantum wave function into a corresponding KvN classical wave function rather than a classical phase space distribution. Since the vectors $| p \, x \rangle$ are identical in both KvN and Wigner representations, $W(x,p)$ is proportional to the probability amplitude that a quantum particle is located at a point $(x, p)$ of the \emph{classical} phase space. Note it is important to distinguish the classical $(\hat{x}, \hat{p})$ and quantum $(\hat{x}_q, \hat{p}_q)$ phase spaces because the notion of a phase space point arises naturally only in the commutative classical variables $(\hat{x}, \hat{p})$. One may take this distinction further and interpret the Wigner function as the KvN wave function of a classical counterpart of the analogous quantum system. Like any wave function, the Wigner function need not be positive. 

\section{A generalization to mixed states}

Equation (\ref{Derivation_Equivalence_of_Averages}) offers a method to extend the developed formalism to an arbitrary state. In particular, we find a fixed vector  $| 1  \rangle \in \mathcal{H}_{xp}$ such that the following equation is valid for all density matrices $\hat{\rho}_{\kappa}$,
\begin{align}\label{Generalization_Equivalence_of_Averages}
	\langle 1 | G(\hat{x}_q) F(\hat{p}_q) | \Psi_{\kappa} \rangle = {\rm Tr}\, \left[ G(\hat{x}_q) F(\hat{p}_q) \hat{\rho}_{\kappa} \right], \quad \forall \hat{\rho}_{\kappa},
\end{align}
where the trace on the right hand side is calculated over the space $\mathcal{H}_{q}$. Let $\{ || \phi_n \rangle\rangle \} $ be a basis in $\mathcal{H}_{q}$ such that $\phi_n(u) = \langle\langle x_q = u || \phi_n \rangle\rangle$. Substituting $\hat{\rho}_{\kappa} = \sum_{n,m} \rho_{n,m} || \phi_n \rangle\rangle \langle\langle \phi_m || $ and $\langle 1 | x \lambda_p \rangle = \sqrt{\hbar\kappa}\chi(u, v)$ into Eq. (\ref{Generalization_Equivalence_of_Averages}) and following the steps in Eq. (\ref{Derivation_Equivalence_of_Averages}), we obtain the equation for the unknown $\chi$
\begin{align}
	\int \chi(u, v) \phi_n^*(v)  dv = \phi_n^*(u),
\end{align} 
which has a unique solution $\chi(u, v) = \delta(u-v)$; therefore,
\begin{align}
	\langle  x \lambda_p | 1 \rangle = \frac{\delta(\lambda_p)}{\sqrt{\hbar\kappa}}, \qquad
	| 1 \rangle = \int \frac{dx}{\sqrt{\hbar\kappa}} \, | x \, \lambda_p = 0 \rangle.
\end{align}
Note that according to Eq. (\ref{UnifiedWaveFunc_connected_Wigner}), the vector $| 1 \rangle$ corresponds to an identity density matrix. 

Equation (\ref{Generalization_Equivalence_of_Averages}) is as a generalization of Eq. (\ref{Equivalence_of_Averages}) to the case when $| \Psi_{\kappa} \rangle$ represents the Wigner function of arbitrary mixed states [see Eq. (\ref{UnifiedWaveFunc_connected_Wigner})].

Additionally, the ket-vector $| 1 \rangle$ maps an observable $F(\hat{x}_q, \hat{p}_q) \in \mathcal{L}(\mathcal{H}_{xp})$, which is an element of the Hilbert space $\mathcal{L}(\mathcal{H}_{xp})$ of linear operators acting on $\mathcal{H}_{xp}$, into  a vector from $\mathcal{H}_{xp}$ as
\begin{align}\label{map_observable_into_kets}
	F(\hat{x}_q, \hat{p}_q) | 1 \rangle  = | F(x_q, p_q) \rangle \in \mathcal{H}_{xp},
	\quad F(\hat{x}_q, \hat{p}_q) \in \mathcal{L}(\mathcal{H}_{xp}).
\end{align}
In other words, each observable corresponds to a vector in the Hilbert phase space, such that the observable's average
 is calculated as the scalar product $\langle F(x_q, p_q) | \Psi_{\kappa} \rangle$  (\ref{Generalization_Equivalence_of_Averages}). 

\section{A realization of the Hilbert phase space $\mathcal{H}_{xp}$}\label{Sec:Hilbert_phase_space_realization}

Both $\mathcal{H}_{xp}$ and $\mathcal{H}_{q}$ have been considered so far as abstract infinite-dimensional vector spaces with no direct connection between them. Following Ref. \cite{Mukunda1978}, we shall construct a realization of the Hilbert phase space $\mathcal{H}_{xp}$ in terms of the space of quantum-mechanical wave functions $\mathcal{H}_{q}$. 

A set of linear operators acting on $\mathcal{H}_{q}$, endowed with the Hilbert-Schmidt inner product $(\hat{A}, \hat{B}) = {\rm Tr}\, (\hat{A}^{\dagger}B)$, forms the Hilbert space $\mathcal{L}(\mathcal{H}_{q})$. Throughout this section, we assume that  $\hat{A}, \hat{B}, \hat{C} \in \mathcal{L}(\mathcal{H}_{q})$. In particular, the set $\mathcal{L}(\mathcal{H}_{q})$ includes all density matrices and observables of the form $F(\hat{x}_q, \hat{p}_q)$.

Define linear operators (i.e., super operators) $\omega(\hat{B})$ and $\Omega(\hat{C})$ acting in $\mathcal{L}(\mathcal{H}_{q})$ as \cite{Mukunda1978}
\begin{align}
	\omega(\hat{B}) \hat{A} = (1/2)\{ \hat{B}, \hat{A} \}, \qquad \Omega(\hat{C}) \hat{A} = [\hat{C}, \hat{A}],
\end{align}
where $\{ \hat{B}, \hat{A} \} = \hat{B}\hat{A} + \hat{A}\hat{B}$ and $\omega(\hat{B}) \hat{A}$ denotes the result of the action of the linear map $\omega(\hat{B})$ on $\hat{A}$ as an element of  $\mathcal{L}(\mathcal{H}_{q})$; a similar interpretation holds for the notation $\Omega(\hat{C}) \hat{A}$.   One can readily demonstrate that
\begin{align}
	& [ \Omega(\hat{B}), \Omega(\hat{C})] \hat{A} = 4 [ \omega(\hat{B}), \omega(\hat{C})] \hat{A} = [[\hat{B},\hat{C}], \hat{A}], \notag\\
	& [\omega(\hat{B}), \Omega(\hat{C})] \hat{A} = (1/2) \{ \hat{A}, [\hat{B}, \hat{C}] \}.
\end{align}
Whence, the Hilbert phase space $\mathcal{H}_{xp}$ can be equated to $\mathcal{L}(\mathcal{H}_{q})$ with the classical operators (\ref{Complete_classical_algebra}) realized as
\begin{align}\label{realization_classical_algebra}
	& \hat{x} = \omega(\hat{x}_q), \qquad \hat{\lambda}_x = \Omega(\hat{p}_q) / (\hbar\kappa) ,  \notag\\
	& \hat{p} = \omega(\hat{p}_q), \qquad \hat{\lambda}_p = - \Omega(\hat{x}_q) / (\hbar\kappa),
\end{align}
where $\hat{x}_q, \hat{p}_q \in \mathcal{L}(\mathcal{H}_{q})$. Other realizations of $\mathcal{H}_{xp}$ can be similarly constructed.

Equations (\ref{Generalization_Equivalence_of_Averages}), (\ref{map_observable_into_kets}), and (\ref{realization_classical_algebra}) bridge the developed Hilbert phase space formalism with the Mukunda approach to the Wigner function \cite{Mukunda1978}. 

\section{Implications for phase space formulation of quantum mechanics}\label{Sec:Commenst_on_phase_space_formalizm}

The Hilbert phase space is an alternative formulation of quantum mechanics obtained as the merger of the wave function and phase space representations via the operational dynamical modeling (figure \ref{Fig_EhrenfestQuantization}). While the connection with the wave-function formulation has already been elucidated in previous sections [in particular, see Eqs. (\ref{Normalization_of_Psi_kappa}), (\ref{Equivalence_of_Averages}), and (\ref{Generalization_Equivalence_of_Averages})], there is more to be considered [in addition to Eq. (\ref{UnifiedWaveFunc_connected_Wigner})] regarding the connection with quantum mechanics in phase space \cite{Bolivar2004, Zachos2005, Curtright2011}.     

The Moyal bracket $\{\{ \,  , \, \}\}$ -- a cornerstone of the quantum mechanical phase space formalism -- is defined as \cite{Curtright1998, Curtright2001, Zachos2005, Curtright2011}
\begin{align}
	\{\{ f , g  \}\} = \frac{2}{\hbar\kappa} f(x,p) \sin\left(
  \frac{\hbar\kappa}{2} \overleftarrow{\frac{\partial}{\partial x}} \overrightarrow{\frac{\partial}{\partial p}} - 
   \frac{\hbar\kappa}{2} \overleftarrow{\frac{\partial}{\partial p}} \overrightarrow{\frac{\partial}{\partial x}}
  \right) g(x,p),
\end{align}
for any smooth functions $f(x,p)$ and $g(x,p)$. Using the identities $ \exp(a \overrightarrow{\partial / \partial y}) f(y) =  f(y) \exp(a \overleftarrow{\partial / \partial y}) = f(y + a)$, we expand the Moyal bracket
\begin{widetext}
\begin{align}\label{MoyalBracketExpantion}
	\{\{ f , g  \}\} &= \frac{1}{i\hbar\kappa} f(x,p) \Bigg[ \exp\left(
  \frac{i\hbar\kappa}{2} \overleftarrow{\frac{\partial}{\partial x}} \overrightarrow{\frac{\partial}{\partial p}}  - 
   \frac{i\hbar\kappa}{2} \overleftarrow{\frac{\partial}{\partial p}} \overrightarrow{\frac{\partial}{\partial x}}
  \right)  - \exp\left(
  -\frac{i\hbar\kappa}{2} \overleftarrow{\frac{\partial}{\partial x}} \overrightarrow{\frac{\partial}{\partial p}} + 
   \frac{i\hbar\kappa}{2} \overleftarrow{\frac{\partial}{\partial p}} \overrightarrow{\frac{\partial}{\partial x}}
  \right)
\Bigg] g(x,p) \notag\\
  &=\frac{1}{i\hbar\kappa} \Bigg[
   f\left(x + \frac{i\hbar\kappa}{2} \frac{\partial}{\partial p}, p - \frac{i\hbar\kappa}{2}\frac{\partial}{\partial x} \right) 
    - f\left(x -\frac{i\hbar\kappa}{2} \frac{\partial}{\partial p} ,p + \frac{i\hbar\kappa}{2} \frac{\partial}{\partial x} \right) 
  \Bigg] g(x,p) \notag\\
  & =\frac{1}{i\hbar\kappa} \langle p \, x | \Bigg[ f\left( \hat{x} - \frac{\hbar\kappa}{2} \hat{\lambda}_p, \hat{p} + \frac{\hbar\kappa}{2}\hat{\lambda}_x \right) - f\left(\hat{x} + \frac{\hbar\kappa}{2} \hat{\lambda}_p ,\hat{p} - \frac{\hbar\kappa}{2} \hat{\lambda}_x \right) \Bigg] | g \rangle,
\end{align}
\end{widetext}
where $g(x,p) = \langle p \, x | g \rangle$. Thus, we established the Moyal bracket realization in the Hilbert phase space. The function $f$ of non-commutative variables in equation (\ref{MoyalBracketExpantion}) should be defined according to the Weyl calculus, as explained in Ref. \cite{Hillery1984}. The operators $\pm\hat{\lambda}_x$ and $\pm\hat{\lambda}_p$ (known as the Bopp operators \cite{Bopp1956, Hillery1984}) are analogues of the left and right derivatives.

An elementary consequence of  Eq. (\ref{MoyalBracketExpantion}) is the equivalence between Eq. (\ref{Dynamical_Eq_for_Wigner}) and Moyal's equation of motion \cite{Zachos2005} (see also Ref. \cite{Dragoman2000}),
\begin{align}
  \frac{ \partial W }{\partial t} = \{\{ H , W \}\}, \quad H(x,p) = \frac{p^2}{2m} + U(x).\label{MoyalEq}
\end{align}  
Here $W$ is the Wigner function of an arbitrary (in general, mixed) state.  
 
\section{Conclusions}\label{Sec:Conclusions}

We demonstrate that the Wigner distribution of a pure quantum state is a wave function [Eqs. (\ref{Dynamical_Eq_for_Wigner}), (\ref{Normalization_of_Psi_kappa}), (\ref{Equivalence_of_Averages})] in the introduced Hilbert phase space. The latter is an alternative formulation of quantum mechanics obtained as a meld of the wave function (the Dirac bra-ket) and phase space (the Wigner function) representations as well as the Koopman von-Neumann classical mechanics. The dynamical equation (\ref{Dynamical_Eq_for_Wigner}) is valid for any non-pure state, providing a starting point for an efficient numerical algorithm for the Wigner function propagation \cite{Cabrera2013}. In the current paper, we have studied dynamics in the cartesian coordinates. In the next work, we will extend our formalism to rotational dynamics.

\acknowledgments
The authors thank anonymous referees for their crucial comments. In particular, we are grateful for drawing our attention to Ref. \cite{Mukunda1978}. The authors are financially supported by NSF and ARO.

\bibliography{literature}
\end{document}